\documentclass[prb,reprint,showpacs,floatfix,superscriptaddress]{revtex4-1}

\usepackage{graphicx}
\usepackage[utf8]{inputenc}
\usepackage[T1]{fontenc}
\usepackage{amsmath}
\usepackage{amsfonts}
\usepackage{amssymb}
\usepackage{xcolor,colortbl}
\usepackage{dcolumn}
\usepackage{multirow}
\definecolor{grey}{RGB}{220,220,220}
\definecolor{matt}{RGB}{192,225,215}

\newcommand{\etc}{etc.}
\newcommand{\etal}{et.\ al.}
\newcommand{\ws}{WS$_2$}
\newcommand{\tis}{TiS$_2$}
\newcommand{\tise}{TiSe$_2$}
\newcommand{\zrs}{ZrS$_2$}
\newcommand{\zrse}{ZrSe$_2$}
\newcommand{\wse}{WSe$_2$}
\newcommand{\mos}{MoS$_2$}
\newcommand{\mose}{MoSe$_2$}
\newcommand{\mosws}{WS$_2$/MoS$_2$} 
\newcommand{\moswse}{WSe$_2$/MoS$_2$} 
\newcommand{\mosews}{WS$_2$/MoSe$_2$} 
\newcommand{\mosmose}{MoS$_2$/MoSe$_2$} 
\newcommand{\mosewse}{MoSe$_2$/WSe$_2$} 
\newcommand{\wswse}{WS$_2$/WSe$_2$}
\newcommand{\tistise}{TiS$_2$/TiSe$_2$}
\newcommand{\tiszrs}{TiS$_2$/ZrS$_2$}
\newcommand{\tisezrse}{TiSe$_2$/ZrSe$_2$}
\newcommand{\abinit}{{\sc{Abinit}\,}}
\newcommand{\opium}{{\sc{OPIUM}\,}}
\newcommand{\oncvpsp}{{\sc{ONCVPSP}\,}}

\newcolumntype{d}[1]{D{.}{.}{#1}}
\begin{document}

\title{Spontaneous interlayer compression in commensurately stacked 
van der Waals heterostructures}
\author{Nicholas A. Pike}
\email[]{Nicholas.pike@ulg.ac.be}
\thanks{Co-first author}
\affiliation{nanomat/Q-MAT/CESAM and European Theoretical Spectroscopy Facility, Universit\'e de Li\`ege, B-4000 Li\`ege, Belgium}

\author{Antoine Dewandre}
\thanks{Co-first author}
\affiliation{nanomat/Q-MAT/CESAM and European Theoretical Spectroscopy Facility, Universit\'e de Li\`ege, B-4000 Li\`ege, Belgium}

\author{François Chaltin}
\affiliation{Department of Chemical Engineering, Universit{\'e} de Li{\`e}ge (B6a), B-4000 Li{\`e}ge, Belgium.}

\author{Laura Garcia}
\affiliation{Department of Physics, Universit{\'e} de Li{\`e}ge (B5), B-4000 Li{\`e}ge, Belgium.}

\author{Salvatore Pillitteri}
\affiliation{Department of Physics, Universit{\'e} de Li{\`e}ge (B5), B-4000 Li{\`e}ge, Belgium.}

\author{Thomas Ratz}
\affiliation{Solid State Physics, Interfaces and Nanostructures/Q-MAT/CESAM, Universit\'e de Li\`ege, B-4000 Li\`ege, Belgium}

\author{Matthieu J. Verstraete}
\affiliation{nanomat/Q-MAT/CESAM and European Theoretical Spectroscopy Facility, Universit\'e de Li\`ege, B-4000 Li\`ege, Belgium}
\affiliation{Catalan Institute of Nanoscience and Nanotechnology (ICN2), Campus UAB, Bellaterra, 08193 Barcelona, Spain.}

\date{\today}
\begin{abstract}
Interest in layered two dimensional materials, particularly stacked heterostructures of transition metal dichalcogenides, has led to the need for a better understanding of the structural and electronic changes induced by stacking. Here, we investigate the effects of idealized heterostructuring, with periodic commensurate stacking, on the structural, electronic, and vibrational properties, when compared to the counterpart bulk transition metal dichalcogenide. We find that in heterostructures with dissimilar chalcogen species there is a strong compression of the inter-layer spacing, compared to the bulk compounds. This compression of the heterostructure is caused by an increase in the strength of the induced polarization interaction between the layers, but not a full charge transfer. We argue that this effect is real, not due to the imposed commensurability, and should be observable in heterostructures combining different chalcogens. Interestingly, we find that incommensurate stacking of Ti-based dichalcogenides leads to the stabilization of the charge density wave phonon mode, which is unstable in the 1T phase at low temperature. Mixed Ti- and Zr- heterostructures are still unstable, but with a charge density wave or a ferroelectric instability.
\end{abstract}


\maketitle

\section{Introduction}
The idea that an ordered stacking of two-dimensional (2D) materials leads to exotic electronic properties has been around since the discovery of graphene~\cite{Novoselov2004}. 2D materials are often used as building-blocks for "materials-by-design"~\cite{Novoselov2016}: each layer displays characteristics that depend on a number of controllable factors including the chemical composition~\cite{pike_2018, Pike2019}, the number of layers, and the rotation angle between individual layers~\cite{Wilson2017, Wang2019}. With the increasing interest in heterostructure devices, it is important to understand, from a first-principles approach, how their structural, electronic, and vibrational properties change due to stacking. This is especially true as experimental limitations are overcome in producing large-area, transferable, monolayers (ML) of material~\cite{Jawaid2017} and heterostructure devices~\cite{Iannaccone2018}.  

\begin{figure*}[ht!]
    \centering
    \includegraphics[width=0.95\linewidth]{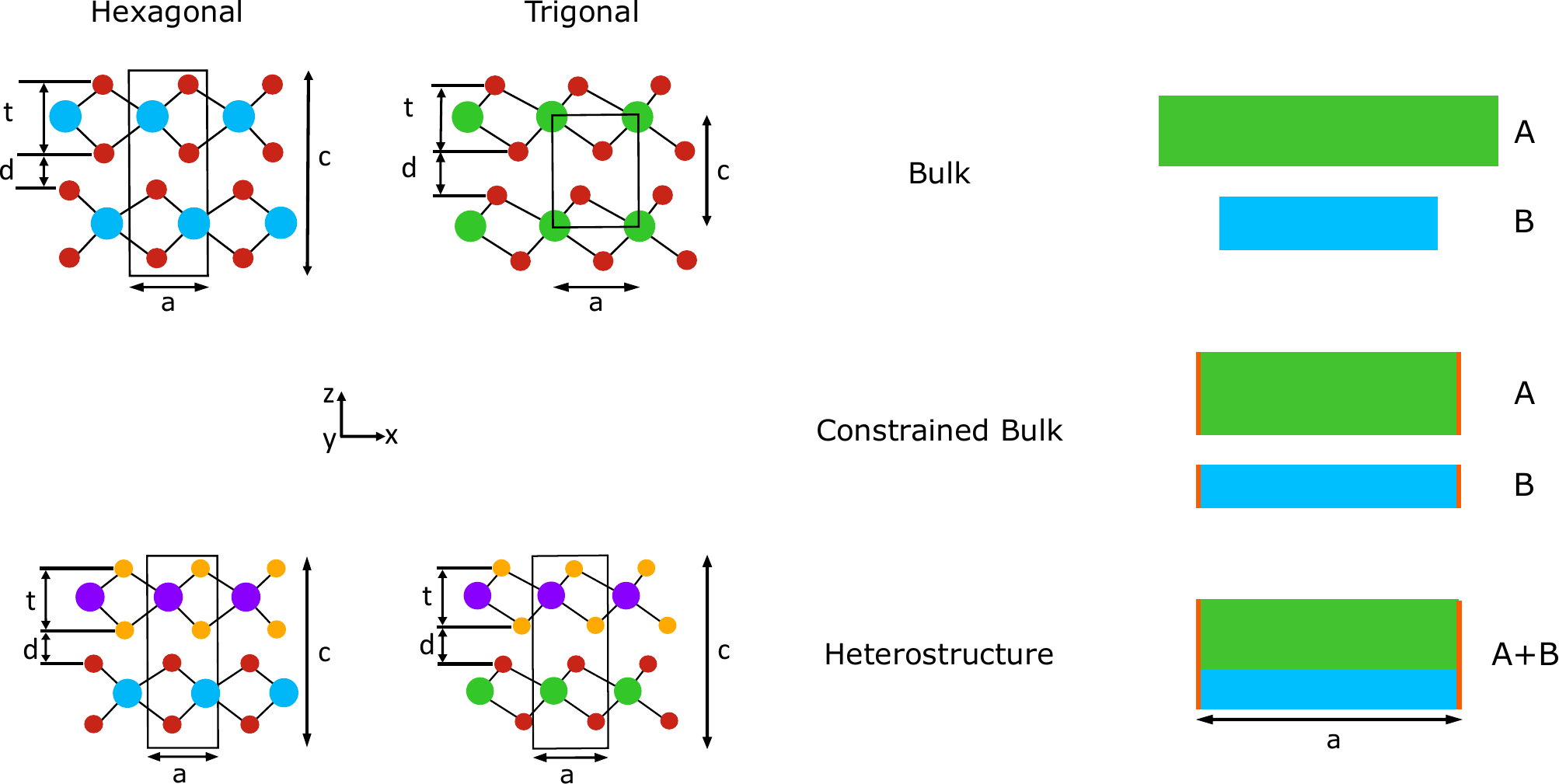}
    \caption{(Color Online) Upper left: Sketches of bulk h-TMDs and t-TMDs with the transition metal atoms as blue/green spheres, chalcogen atoms in red spheres, and unit cell coordinate system (middle left). The parallelograms in black represent the unit cells for each structure. The difference in stacking sequences for the TMDs can be observed in the lower left of the figure. H-TMDs have AB stacking in the $z$ direction while the t-TMDs have AA stacking in the $z$ direction. The geometric thickness ($t$) and vdW gap ($d$) is shown in relation to the out-of-plane lattice constant ($c$) for all cases. Right side: Cartoon representation of our calculation process. Given two bulk structures, A and B in the upper right, we generate the heterostructure (A+B, lower right) and use the relaxed lattice parameters of A+B to generate the constrained bulk structures (middle right). Orange vertical bars indicate that the in-plane lattice parameters of the heterostructure and constrained calculations are identical.  }
    \label{figure: structures_bulks}
\end{figure*}

Experiments combine (preferably air-stable) ML materials (graphene, black phosphorus, hexagonal boron-nitride, \mos, \etc) into heterostructure devices, either by creating lateral junctions via in-plane covalent bonding~\cite{wei2015} or by growing or stacking them vertically~\cite{Chen2019}. These devices are used as biosensors~\cite{Bolotsky2019}, field-effect transistors~\cite{Zhou2019}, photo-detectors~\cite{li_2016}, electro-optical devices~\cite{Wang2018b, rivera2015, Taghinejad2019, Lattyak2019}. Cao~\etal~\cite{Cao2018} even showed that at very small angles twisted bilayer graphene can become superconducting. Stacked heterostructures with different transition metal dichalcogenides (TMDs) (e.g. \mose~on \wse) have long-lived inter-layer excitons, resulting in spatially separated electrons and holes~\cite{rivera2015, Wang2018, Torun2018}, an ideal feature for photovoltaic applications. Stacking of non-TMD layered materials into incommensurate heterostructures has also been shown to lead to enhancements in the electronic specific heat of the heterostructure and stabilization of native charge density waves~\cite{bao2020}. Low frequency Raman measurements of heterostructure materials~\cite{Lui2015,Amin2016, Zhang2016c} can be used to determine the strength of layer interactions in a heterostructure lattice. The presence and frequency shifts of in-plane shear modes show commensurate stacking and strain effects. Even for incommensurate materials, the interlayer breathing mode is always present and its frequency and width reveal details of the layer distance and bonding strength.

Recent high-throughput calculations, see for example Refs.~\citenum{mounet_2018, 2017_choudhary_scirep, Zhang2019} and~\citenum{Curtarolo2012}, have discovered numerous two-dimensional materials that could be used to create devices with unique electric and optical properties. The combinatorial phase space for 2D heterostructures is immense and there are many unique and exotic phenomena yet to be explored both experimentally and theoretically. Ab initio simulations determining the properties of stacked heterostructures have expanded significantly in the past few years, but are often limited to commensurate cases, despite the large sizes of the super-lattices that are observed experimentally.

Recent work by Van Troeye~\etal~\cite{troeye2019} on van der Waals (vdW) heterostructures used first-principles ingredients to predict the coherence of a heterostructure from the individual ML elastic constants and structural parameters. Work by Pizzi~\etal~\cite{pizzi2020} on layer-dependent interactions in stacked materials composed of the same ML used symmetry arguments and a spring model to understand layer dependent vibrational properties in vdW structures. In the hexagonal TMDs, Phillips~\etal~\cite{phillips2019} and Terrones~\etal~\cite{Terrones2013} calculated the electronic structure of experimentally realized bilayer heterostructures in both AA and AB stacking assuming a commensurate structure and found both direct and indirect electronic band gaps.

In the following, we investigate heterostructures of vertically stacked TMD materials using first-principles calculations. This approach ensures a consistent interpretation of the results across a wide variety of physical properties. In section~\ref{sec: crystal structure}, we describe the structure of our TMD stacks, for the two crystal symmetries used here. Section~\ref{sec: calculation methods} describes our first-principles methods, and Section~\ref{sec: results} details and discusses the results of our calculations for all the commensurate heterostructures, in particular the interlayer compression. Finally, in Section~\ref{sec: conclusions}, we provide concluding remarks. 

\section{Crystal Structure}\label{sec: crystal structure}
In their bulk forms, the TMD compounds are stacked layers bound by vdW forces. They belong to two main symmetry classes: hexagonal (P6$_3$/mmc, space group 194) and trigonal (P$\overline{3}$m1, space group 164). A few metals, such as Re, produce a closely related triclinic structures (P$\overline{1}$, space group 2). The structures are similar, with layers of trigonal or octrahedral prisms, but give different stacking orders in the bulk. An example of these two materials is shown in Fig.~\ref{figure: structures_bulks} in which the bulk structures of generic hexagonal and trigonal TMDs are shown on the left and the commensurately stacked heterostructures are on the right. The hexagonal TMDs have AB stacking, as a result of the 180$^\circ$ rotation between the layers, and the trigonal TMDs have AA stacking. We denote the structures here as, e.g. \mosewse, in the notation of Ref.~\citenum{Tritsaris2020}, omitting the rotation angle between the layers. In our case this angle is always 0, but in reality many intermediate angles can be chosen, resulting in long range moir\'e patterns~\cite{Kang2013, Wijk2015}.

Our heterostructures are composed of two different TMD layers repeating periodically in the out-of-plane direction. The number of possible combinations is restricted by studying only combinations of compounds with the same symmetry.  It is important to note that the hexagonal TMD MLs do not possess inversion symmetry (only the bulk crystal does), whereas trigonal TMD MLs do, and this carries over to the heterostructures they compose. Thus, for hexagonal MLs stacked as a heterostructure, the space group of the heterostructure becomes P$\overline{6}$m2 (space group 187).

\begin{figure*}[th!]
    \centering
        \includegraphics[width=\linewidth]{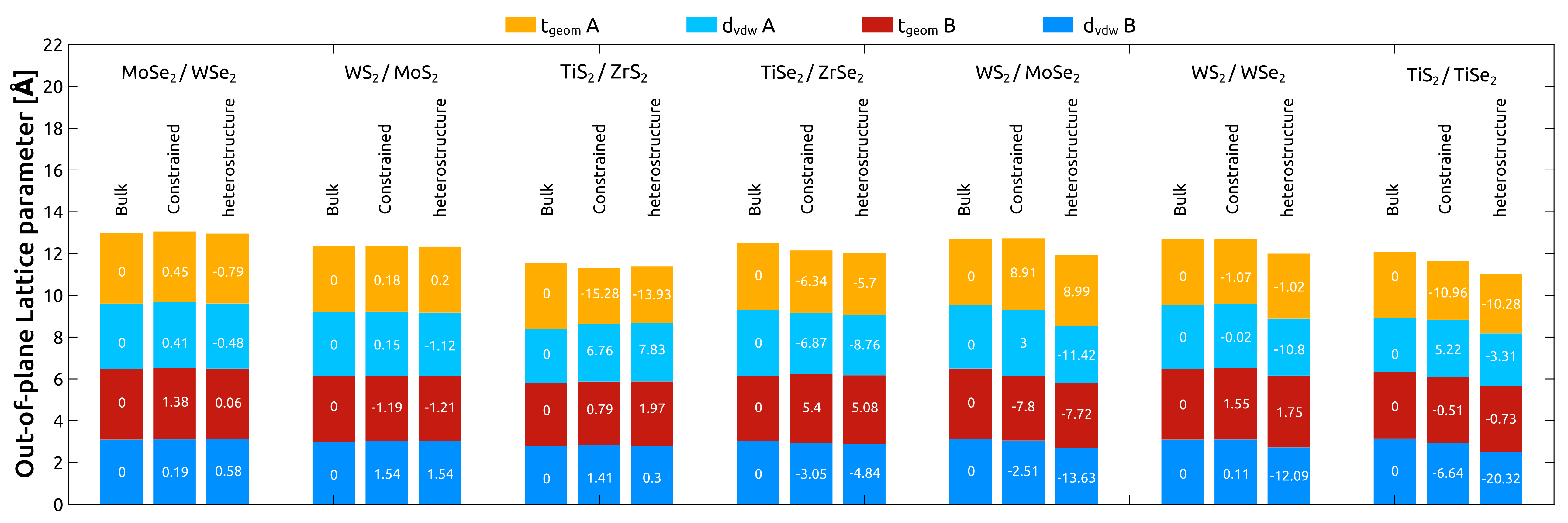}
    \caption{Calculated out-of-plane lattice parameter (full height of each bar) for each of the heterostructures considered here. Each bar is broken down into contributions from the geometric thickness ($t$) and vdW gap ($d$) of each ML as described in Figure~\ref{figure: structures_bulks}. For each set of three bars, Left: sum of the bulk values of $t$ and $d$ (Ref.~\citenum{pike_2018}). Right: fully relaxed heterostructures. Middle bar: sum of constrained calculations, for each material with in-plane lattice matching the relaxed heterostructure.  
    The inset white numbers correspond to the amount of compression or expansion (in percent) of the individual thickness/spacing, when compared to bulk values. A and B correspond to the first and second material given in the title of each triplet of bars. There is a systematic trend to global compression, which is much stronger (up to 10\% and more) in the mixed chalcogen cases (group II, right 3 sets).}
    \label{figure: structures_all}
\end{figure*}

\section{Calculation Methods}\label{sec: calculation methods}
We employ density functional theory (DFT)~\cite{Martin2004} and density functional perturbation theory (DFPT)~\cite{Gonze1997b,Baroni2001,Gonze1997a} calculations, as implemented in the \abinit~software package~\cite{Gonze2005,Gonze2009,Gonze2016,Romero2020}, with a plane wave basis set and norm-conserving pseudopotentials. The pseudopotentials used here are generated use the fhi98pp code~\cite{Fuchs1999}, except for W and Ti which produced inaccurate lattice parameters (errors over 5\%) or overly large band gap energies.  For W, we use a pseudopotential generated with the \opium code~\cite{opium_psp} and for Ti the pseudopotential is generated using the \oncvpsp software~\cite{Hamann2013}. In all cases, the pseudopotentials use the GGA-PBE exchange-correlation functional~\cite{Perdew1996} without spin-orbit coupling. As shown in Ref.~\citenum{pike_2018} the inclusion of spin-orbit interactions is not necessary in these TMD compounds for the structural and vibrational properties. 

Our calculations of the heterostructures follow careful convergence studies of the energy cut-off (E$_{cut}$) and of the reciprocal space $k$-point mesh for each of the compounds. A plane wave cutoff energy of E$_{cut}=$ 50 Ha and a k-point grid size of $8 \times 8 \times 8$ ($16 \times 16 \times 8$ for the metallic TMD systems) guarantee total energies to within 3 meV per six atom unit cell for all the compounds. The self-consistent cycle is converged to total energy differences less than 10$^{-10}$Ha. For the relaxation of the structural parameters we use the Broyden-Fletcher-Goldfarb-Shanno minimization procedure~\cite{Fletcher_BFGS} in which we allow both the position of the atoms and cell shape to change simultaneously, while imposing symmetry. Convergence of this relaxation procedure stopped when the forces are lower than 1 $\mu$Ha/Bohr. Our DFPT calculations for the inter-atomic force constants used the same optimized $E_{cut}$ and the relaxed geometries. Static atomic charges are determined from a Bader atom-in-molecule approach~\cite{Bader1985} within~\abinit.

\begin{table}[bh!]
    \centering
    \begin{tabular}{|c| d{1.3}| } \hline
    Bulk & \multicolumn{1}{c|}{E$_g$(eV)}\\ \hline
 \mos    & 0.886   \\
 \mose  & 0.853   \\
 \ws    & 1.029   \\
 \wse   & 0.938   \\
 \tis   & \multicolumn{1}{c|}{-{}-}  \\
 \tise  &  \multicolumn{1}{c|}{-{}-} \\ 
 \zrs   & 1.041 \\
 \zrse  & 1.004 \\ \hline
 \multicolumn{2}{c}{}\\ 
 \multicolumn{2}{c}{}\\
 \multicolumn{2}{c}{}\\
 \multicolumn{2}{c}{}
\end{tabular}
    \begin{tabular}{|c| d{1.3} d{1.4} | } \hline
Heterostructure & \multicolumn{1}{c}{E$_{avg}$(eV)}&\multicolumn{1}{c|}{E$_g$(eV)} \\ \hline
\mosewse &0.895 & 0.879 \\
 &      & 1.12 \footnote{Ref.~\citenum{Terrones2013} Theory}\\
\mosws   & 0.958 & 1.049 \\
   &      & 1.23\footnote{Ref.~\citenum{phillips2019} Theory}\\
\tiszrs  &  \multicolumn{1}{c}{-{}-} & 0.766 \\
\tisezrse & \multicolumn{1}{c}{-{}-} & 0.159 \\ \hline
\mosews  &0.941 & 0.722 \\
 &      & 1.05\footnotemark[1]\\
 &      & 1.58 \footnote{Ref.~\citenum{Amin2016} Theory}\\
\wswse   &0.983 & 0.817 \\
 &      & 1.07\footnotemark[1]\\
\tistise & \multicolumn{1}{c}{-{}-} & 0.059 \\ \hline
    \end{tabular}
    \caption{Calculated electronic band gap energies for each of our heterostructures and their bulk counterparts. For the heterostructures, we also report the average of the host material band gaps. -{}- indicates that the calculated structure is metallic. NB: The calculated values of Refs.~\citenum{phillips2019} and~\citenum{Terrones2013} are for commensurate bilayers, not periodic heterostructures.}
    \label{table: electronic gaps}
\end{table}

To take into account the long-range electron-electron correlation, we use the dispersion scheme given by Grimme~\cite{Grimme2010} which is known to reproduce the interactions of 2D materials quite well~\cite{Troeye2016, pike_2018, Sabatini2016}. This dispersion scheme, known as DFT-D, is based on simple atomic pair-wise terms, with environment-dependent dispersion coefficients tabulated beforehand using time-dependent DFT. We showed previously that DFT-D3 in particular is very accurate compared to experiment in the 2D TMDs~\cite{pike_2018}. More generally, the performance of the most-popular vdW functional and dispersion schemes have been assessed recently~\cite{Tawfik2018} and DFT-D3 performs well compared to more elaborate methods.

For all the vibrational property calculations we numerically converge the ground state wave function to a relative variance less than $10^{-18}$, and the first order wave functions to less than $10^{-10}$. The dynamical matrices are calculated on an irreducible Brillouin Zone wedge of $q$-points corresponding to an unshifted $4 \times 4\times 4$ mesh. 

The overall calculation procedure for each heterostructure is as follows: First, we generate and fully relax the periodically stacked commensurate heterostructures A/B (third column of data in figure~\ref{figure: structures_all}). These heterostructure geometries differ from the bulk due to two main effects: the juxtaposition of the layers (both short-range ``chemical'' and long-range dielectric interaction), and the strain imposed by the artificial commensuration. Second, to separate these effects, we generate constrained bulk structures for each component, e.g. A, constraining the in-plane lattice parameters of a bulk unit cell of A to those of the heterostructure, and allowing the out-of-plane lattice parameter to relax (middle bars in figure~\ref{figure: structures_all}). After relaxation we calculate the electronic and vibrational properties of each system.

\section{Results}\label{sec: results}
Comparing the heterostructure to isolated MLs, the relaxation of a multi-layered system must account for the lattice mismatch between the individual crystalline layers. This mismatch adds a strain energy to the interface, which generically will compress one layer and expand the other. In reality, 2D vdW systems often release this strain energy by rotating one layer with respect to the other, forming commensurate long range relaxation patterns (moir\'e) or incommensurate interfaces~\cite{troeye2019}. Given that our structures are forced to be commensurate, the strain energy leads to an intermediate in-plane lattice parameter, indicated in Table 1 of the supplemental material (SM)~\cite{supmat1}, which integrates the elasticity of each layer (the softer material will accommodate its lattice more). The table also compares the out-of-plane lattice parameters of our heterostructures to their bulk counterparts~\cite{pike_2018} and to measurements of the vdW gap~\cite{Wilson1969}.

\begin{figure*}[th!]
    \centering
    \includegraphics[width=0.49\linewidth]{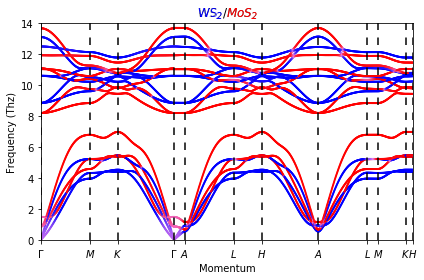}
    \includegraphics[width=0.49\linewidth]{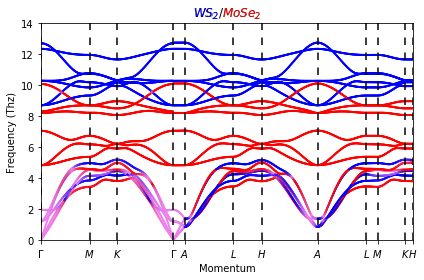}\\
    \includegraphics[width=0.49\linewidth]{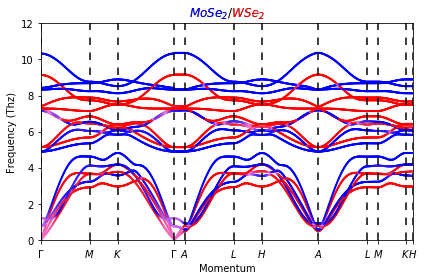}
    \includegraphics[width=0.49\linewidth]{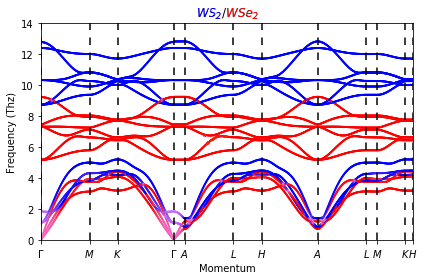}
    \caption{Phonon dispersion relations for the hexagonal heterostructures projected by layer (sum of projected eigendisplacement weight on the atoms in the layer). The color indicates the fraction of the phonon mode on each layer. Blue: 100\% first TMD; Red: 100\% second TMD; Magenta: hybridized mode. The optical modes are very well separated by layer, whereas the acoustic manifold is more mixed, in particular for~\mosews. The high-symmetry path in momentum space used for our calculations is shown in Figure~\ref{figure: phonon trigonal}.}
    \label{figure: phonon hexagonal}
\end{figure*}

Figure~\ref{figure: structures_all} shows a bar graph comparing the out-of-plane lattice parameters in different conditions: sum of the bulk compounds, of the constrained heterostructures, and the fully relaxed heterostructure calculations. In each case, the calculated out-of-plane lattice parameter, $c$ can be subdivided into contributions from each layer's geometric thickness and vdW spacing. For layers A and B, $c=t_{geom, A}+d_{vdW, A}+t_{geom,B}+d_{vdW,B}$ (in the heterostructures the vdW spacing's are equivalent and ``shared'' by both layers).

Comparing the height of the relaxed heterostructure (right bar in each set of figure~\ref{figure: structures_all}) to the sum of the heights of the bulk compounds first relaxed (left column) then epitaxially constrained (middle) we find there are two groups of materials. In group I (first 4 sets in figure~\ref{figure: structures_all}), there is only a small compression of the out-of-plane lattice parameter (maximum 5\%) and most of it is due to strain and the unequal elastic responses of the two materials: for idealized equal and isotropic elastic responses, all of the expansion in A would be a contraction in B and the three bars would be identical.

In group II, there is a strong contraction of the out-of-plane lattice parameter (around 10\% on average), and most of the effect is due to the interaction between the layers (change from middle to right bars). The main difference between the two groups is that, in group I, the chalcogens of the two layers are identical (\mosewse, \mosws, \tiszrs, and \tisezrse) whereas, in group II, the chalcogens are different (\mosews, \wswse, and \tistise). Figure~\ref{figure: structures_all} shows that the change in out-of-plane lattice parameter is mainly located in a contraction of the vdW gap ($d_{vdw}$, blue segments). Changes in the geometric thickness ($t$ red and orange segments), on the other hand, are mainly caused by strain. The reason for the vdW gap contraction in Group II will be explored through several derived properties below.

The compression of the out-of-plane lattice parameter, through the reduction of the vdW gap, leads to important changes in the electronic and vibrational properties of the system. Plots of the Kohn-Sham band structures for each of the compounds are given in SM Figures 1 and 2~\cite{supmat1}. To quantify the change, we compare in Table~\ref{table: electronic gaps} the calculated DFT electronic band gap energies for the heterostructures with the average band gap energy of the heterostructure components (bulk compound references listed as well). We report band gap energies for the constrained structures in SM Table 2~\cite{supmat1}. Interestingly, within group II, the resulting band gap energies for all semiconducting heterostructures are smaller than the average and the individual band gaps of their bulk counterparts, whereas the~\tistise~heterostructure becomes semiconducting. Within group I, the average bulk and heterostructure band gaps are comparable. In all of the hexagonal cases, the $Q$ valley between $\Gamma$ and $K$ is lower (indirect band gap SM Figure 1), as in the bulk TMD compounds. This will strongly affect the transport and optical properties of the stack, but will also be very sensitive to the precise geometry: a bilayer may be very different from a periodic continuous heterostructure, and strain can invert the valley ordering.

\begin{figure*}[th!]
    \centering
    \includegraphics[width=0.49\linewidth]{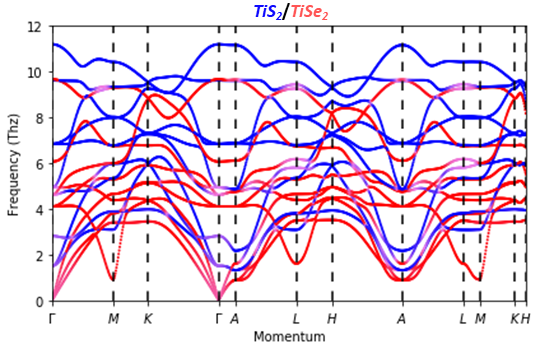}  
    \includegraphics[width=0.49\linewidth]{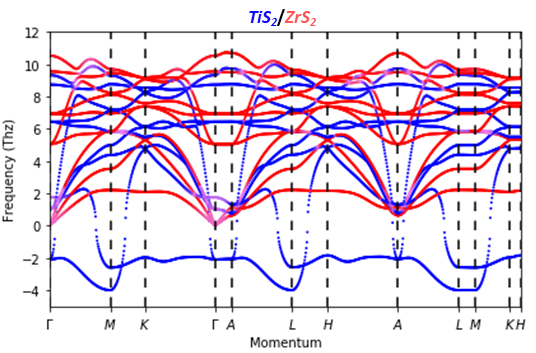}\\
    \includegraphics[width=0.49\linewidth]{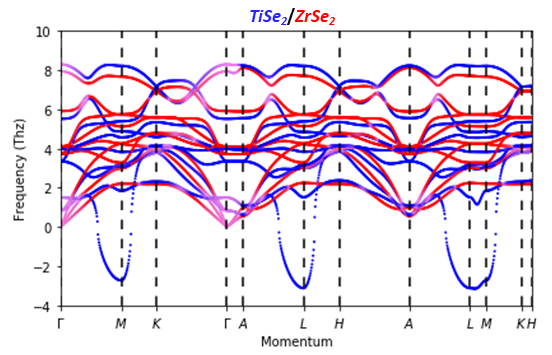}
    \includegraphics[width=0.49\linewidth]{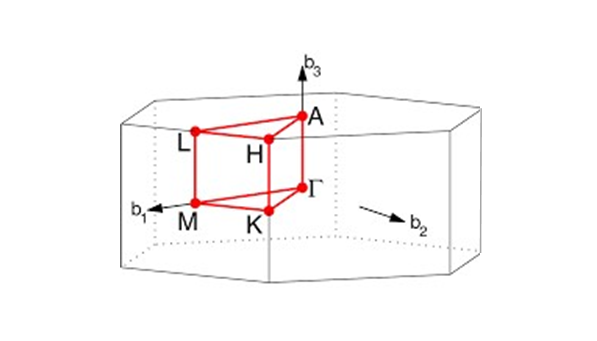}
    \caption{Phonon dispersion relations of the trigonal heterostructures projected by layer (sum of projected eigendisplacement weight on the atoms in the layer). The color indicates the fraction of the phonon mode on each layer. Blue: 100\% first TMD; Red: 100\% second TMD; Magenta: hybridized mode. Lower right corner: High-symmetry path in momentum space used for our band structure plots from Ref.~\citenum{Setyawan2010}.}
    \label{figure: phonon trigonal}
%
\end{figure*}

Given the changes in the lattice parameters of the heterostructures compared to their bulk counterparts, we expect changes in the vibrational structure of these compounds.  We use DFPT to calculate the vibrational properties of our systems, determining the inter-atomic force constants as well as responses to homogeneous electric fields, giving access to the dielectric constant and the Born effective charges (BEC). The inter-atomic force constants allow us to determine the phonon dispersion relations and related properties (e.g. free energies).

\subsection{Vibrational Properties}
In Figures~\ref{figure: phonon hexagonal} and~\ref{figure: phonon trigonal}, we show the phonon dispersion curves of the heterostructures. We determine the contribution from each layer by projecting the phonon mode displacement vectors onto the atoms in each layer, and color the phonon band structure accordingly. In Figure~\ref{figure: phonon hexagonal}, we show the phonon dispersion relations for hexagonal heterostructures \mosws, \mosews, \mosewse, and \wswse. There is a clear separation between the optical modes from each layer, which is a signature of weak vdW bonding between layers. The only noticeable mixing between the layers occurs in certain acoustics modes near $\Gamma$, and avoided crossings which appear with exchanges of layer character (blue/red switching).  In \mosews~(group II), a significantly stronger mixing occurs in the acoustic branches along the in-plane momentum path. Furthermore, the weak interaction between the layers is supported by small dispersion in the out-of-plane direction (segments $\Gamma$-$A$, $K$-$H$ and $L$-$M$).  

A number of studies have examined the phonon modes of TMD bilayers such as ~\citenum{Lui2015, Amin2016, Zhang2016c, Liang2017} (our list is not exhaustive and focuses on works with dissimilar chalcogens). In particular Ref.~\citenum{Lui2015} pioneered the field by tracking the Raman shifts in different hetero-bilayer combinations, including \moswse~and \mosmose. The shear mode disappears in the Raman spectra due to the layer incommensurability (it survives in our calculations due to the enforced epitaxy), but the layer breathing mode is always present, with a frequency between those of the two component bulks, showing strong mode mixing as we see in this work. Table~\ref{Tab: LBM frequency} compares our calculated interlayer breathing mode frequencies to available literature data. Our frequencies are systematically higher, due to the imposed commensuration and to the periodic nature of our structures out of plane (3D periodicity instead of free bilayers), which will increase the effective interlayer spring constant~\cite{Liang2017}. 

\begin{table}[b!]
    \centering
    \begin{tabular}{|c| c| } \hline
    Heterostructure & \multicolumn{1}{c|}{$\omega_{LBM}$(cm$^{-1}$)} \\ \hline
\mosmose   &  $\approx$ 32~\footnote{Ref.~\citenum{Lui2015} experiment on bilayer} \\
\mosewse   & 39.8\\
 \mosws    & 49.3 \\
           & 31.2~\footnote{Ref.~\citenum{Zhang2016c} experiment on bilayer}\\
 \tiszrs   & 33.9\\
 \tisezrse & 50.1\\ \hline
 \mosews   & 64.7\\
           & 39~\footnote{Ref.~\citenum{Amin2016} theory on bilayer}\\
 \moswse   & $\approx$ 32~\footnotemark[1]\\
 \wswse    & 61.1\\
 \tistise  & 94.0\\ \hline
\end{tabular}
\caption{\label{Tab: LBM frequency}Frequencies of the interlayer breathing mode ($\omega_{LBM}$) for our commensurate heterostructures divided into group I and group II structures. Experimental Raman data for two related structures are shown for comparison.}
\end{table}

In Figure~\ref{figure: phonon trigonal}, we present the trigonal heterostructures (\tistise, \tiszrs, and \tisezrse) which show a qualitatively different and more complex behavior. The trigonal heterostructure phonon modes are more strongly hybridized than the hexagonal ones (including some optical modes as well), visible in the more numerous mixed magenta bands. There is no splitting between the acoustic and optical manifolds, which is the case in the bulk as well: the bonding is softer than in hexagonal TMDs. 

Bulk \tis~and \tise~are challenging to simulate because they present a charge-density wave (CDW), leading to a phonon instability that is very sensitive to unit-cell volume, pseudopotential, and strain~\cite{Dolui2016, Guster2018, Chen2015}. Experimentally and theoretically, the phonon instability in \tis~and \tise~appears at the M special point for both bulk and ML compounds.  

For \tistise, we find that stacking dynamically stabilizes the system: the heterostructure stays metallic, but there are no unstable phonon modes. Recent work by Liao~\etal~\cite{Liao2020} shows that, in these CDW materials, induced strain is not enough to stabilize the phonon mode, and that charge transfer is most likely necessary for system stabilization. This is the case for our Ti based heterostructures as well: both layers become doped as shown in Table~\ref{table: born charges}, and there is a net charge transfer between the layers. Stabilization via charge transfer is also thought to occur in non-TMD layered materials~\cite{bao2020}.

We find that \tiszrs~and \tisezrse~are dynamically unstable: in \tisezrse~the system remains semi-metallic, and the phonon dynamics resemble the CDW in \tise~\cite{Guster2018}, with an instability at M. For \tiszrs, the band structure becomes semiconducting, and the layers are roughly charge neutral. Here the stronger instability has a different origin, with a Ti-layer optical branch which is imaginary in the full Brillouin zone. At $\Gamma$ the two lowest modes are polar in-plane displacements of the Ti against S, i.e. ferroelectric. Given the experimental demonstration of CDW suppression in heterostructures~\cite{bao2020}, it would be very interesting if this behavior could be confirmed experimentally as well: 2D ferroelectrics are highly sought after for ultrathin and flexible sensors, actuators and memory devices. The phonon displacement vectors at M for the lowest imaginary mode are in-plane, with the two S atoms moving in opposite directions along a (110) high-symmetry axis. This is similar to the CDW displacements in bulk, and to the \tisezrse~instability at M. For the second unstable mode at M, the eigendisplacements follow the (210) in-plane direction. There is a significant out-of-plane component, and a net movement of the \tise~layer against the \zrse~layer, which produces local Peierls-like dimerization and global buckling of the layers.

\subsection{Origin of the vdW contraction}
A natural explanation for the contraction of the vdW gap would be charge transfer between layers, leading to Coulombic attraction which is not present in the bulk. A strong charge transfer would make the two layers metallic, with a net electron(hole) excess on layer A(B). This is not the case for the semiconducting TMDs, as the electron band structures show no metallization (see SM Figures 1 and 2 for band structure plots~\cite{supmat1}). 

A weaker effect would be charge polarization in each layer. To quantify charge transfer and displacement in our heterostructures, we calculate the dynamical (Born) and static (Bader) charges for each atom, compared to the bulk. The dynamical charges are shown in Table~\ref{table: born charges}, and systematically increase in the heterostructure environment compared to the bulk. We have highlighted the cells in which there are significant differences (> 5\%) with respect to the reported bulk BEC in Ref.~\citenum{pike_2018}, which happens almost systematically in the out of plane direction. We even find that the BEC of WS$_2$ in mixed group II heterostructures become larger out-of-plane than in-plane (yellow cells). In hexagonal systems, the counterintuitive sign and abnormal magnitude of the BEC observed in bulk is still apparent even under compression of the unit cell and heterostructure stacking~\cite{Pike2016}.

For the trigonal compounds, one can not compare to the (metallic) bulk, as the calculated BEC are not meaningful. In semiconducting~\tistise, the large values for Ti atoms are similar to those found, e.g., in oxide  perovskites~\cite{Ghosez1998}. The large BEC are linked to the small electronic gap and the sizable dielectric polarizability.

\begin{table}[t!]
    \centering
    \begin{tabular}{|c | c | d{3.2} d{3.2} | d{3.2} d{3.2}| c |}   \hline
    & atom & \multicolumn{2}{c|}{Z$^*_{xx}$}  & \multicolumn{2}{c|}{Z$^*_{zz}$}  & \multicolumn{1}{c|}{q$_{B,L}$(e)}\\ 
    &   & het & bulk  & het & bulk  & \\ \hline
\multirow{4}{*}{\mosws}& Mo &\cellcolor{grey}-1.18 & -1.09&-0.63  & -0.63 &\multirow{2}{*}{0.003} \\
                       & S & \cellcolor{grey}0.59 &  0.54& \cellcolor{grey}0.33& 0.31 &\\
                       & W & \cellcolor{grey}-0.52& -0.49 & \cellcolor{grey}-0.47&-0.43  &\multirow{2}{*}{0.007} \\
                       & S & 0.26&  0.25 & 0.22 &0.21 &\\ \hline
\multirow{4}{*}{\wswse}& W & -0.48 & -0.49 & \cellcolor{yellow}-0.92 & -0.11 &\multirow{2}{*}{-0.027} \\
                       & S &  0.26 &  0.25 & \cellcolor{yellow} 0.53 &  0.21 & \\
                       & W & \cellcolor{grey}-1.32 & -1.24 & \cellcolor{grey}-0.81 & -0.30 &\multirow{2}{*}{0.035} \\
                       & Se&  0.65 &  0.62 & \cellcolor{grey} 0.33 &  0.39& \\\hline
\multirow{4}{*}{\mosews} & Mo &\cellcolor{grey}-2.02 & -1.91 & \cellcolor{grey}-1.00 & -0.95 &\multirow{2}{*}{0.030} \\
                         & Se &  0.96 & 0.95  & \cellcolor{grey}0.44 &  0.48 & \\
                         & W  & -0.49 &-0.49  &\cellcolor{yellow} -1.00 & -0.11 &\multirow{2}{*}{-0.025} \\
                         & S  &  0.26 & 0.25  &\cellcolor{yellow} 0.54  & 0.21& \\\hline
\multirow{4}{*}{\mosewse} & Mo & -1.99 & -1.91 & -0.99 & -0.95 &\multirow{2}{*}{0.000} \\
                          & Se &  \cellcolor{grey}1.00 & 0.95  & \cellcolor{grey}0.52 & 0.48 & \\
                          & W  & -1.25 &-1.24  &\cellcolor{grey}-0.80 & -0.30 &\multirow{2}{*}{0.008} \\
                          & Se &  0.62 & 0.62  & 0.38 & 0.39 &\\\hline
\multirow{4}{*}{\tistise} & Ti &  \multicolumn{1}{c}{-{}-}&  \multicolumn{1}{c|}{-{}-} &  \multicolumn{1}{c}{-{}-} & \multicolumn{1}{c|}{-{}-} &\multirow{2}{*}{-0.032}\\ 
                          & S  & \multicolumn{1}{c}{-{}-} & \multicolumn{1}{c|}{-{}-} & \multicolumn{1}{c}{-{}-}  &  \multicolumn{1}{c|}{-{}-}&\\ 
                          & Ti & \multicolumn{1}{c}{-{}-} & \multicolumn{1}{c|}{-{}-} &  \multicolumn{1}{c}{-{}-} & \multicolumn{1}{c|}{-{}-} &\multirow{2}{*}{0.030}\\ 
                          & Se & \multicolumn{1}{c}{-{}-} & \multicolumn{1}{c|}{-{}-} & \multicolumn{1}{c}{-{}-}  & \multicolumn{1}{c|}{-{}-} &\\ \hline 
\multirow{4}{*}{\tiszrs}  & Ti & \multicolumn{1}{c}{-{}-} & \multicolumn{1}{c|}{-{}-} & \multicolumn{1}{c}{-{}-}  & \multicolumn{1}{c|}{-{}-}&\multirow{2}{*}{0.002} \\ 
                          & S & \multicolumn{1}{c}{-{}-} & \multicolumn{1}{c|}{-{}-} & \multicolumn{1}{c}{-{}-}  & \multicolumn{1}{c|}{-{}-}&\\ 
                          & Zr & \multicolumn{1}{c}{-{}-} &  6.19 & \multicolumn{1}{c}{-{}-} & 1.82  &\multirow{2}{*}{-0.006}\\ 
                          & S  &\multicolumn{1}{c}{-{}-} & -3.17 &\multicolumn{1}{c}{-{}-}  &-0.60 & \\\hline 
\multirow{4}{*}{\tisezrse} &Ti &  6.63 & \multicolumn{1}{c|}{-{}-}     & 1.03 & \multicolumn{1}{c|}{-{}-}  &\multirow{2}{*}{0.011}\\
                           &Se & -3.24 &  \multicolumn{1}{c|}{-{}-}    & -0.50 & \multicolumn{1}{c|}{-{}-}& \\
                           &Zr & \cellcolor{grey} 7.66 & 6.76  & \cellcolor{grey}1.89 & 1.70  & \multirow{2}{*}{-0.007}\\         
                           &Se & \cellcolor{grey}-3.90 & -4.12 &\cellcolor{grey}-0.94 &-0.55 & \\\hline
    \end{tabular}
    \caption{Comparison of dynamical and static charges in the TMD heterostructures. The calculated BEC for each of the heterostructures are compared to their bulk counterpart, both in-plane (xx) and out-of-plane (zz). The first column for each BEC direction corresponds to the heterostructure and the second column to the bulk. Shaded cells highlight heterostructure values that differ by more than 5\% from the bulk, and yellow cells represent compounds with a larger out-of-plane than an in-plane value of the BEC. -{}- indicates that the calculated structure is metallic. Most of the zz values increase perceptibly, and none of the values decrease: heterostructuring always increases polarizability with respect to a chemically homogeneous reference bulk.
    The last column contains the static Bader charges, summed over a layer (L) (bulk reference is 0). The heterostructures with identical chalcogens have negligible charge transfer, within the numerical noise of the Bader charge integration routine, whereas the hetero-chalcogen cases show transfers of 0.02 to 0.03 electrons.}
    \label{table: born charges}
\end{table}

Our calculated values of the static Bader charges, for both bulk and heterostructure systems, are given in Table 3 of the SM~\cite{supmat1}, and the layer charges (sum of atomic charges for layers A and B), which are shown in the last column of Table \ref{table: born charges}. The charges show small changes with respect to the bulk compounds, but there is a clear difference between the homo-chalcogen cases (few milli-electron charges for the layers) and the hetero-chalcogens (an order of magnitude more), in favor of the sulfide layers.

At this point we have ingredients to understand the shrinking of the interlayer gaps presenting different chalcogens. Changing the bulk environment systematically induces a greater polarizability in both materials, generating additional attractive forces between the layers. The modified charge density gives rise to an induced polarization within the layers, both static (Bader charge imbalance) and dynamic (larger BEC, especially out of plane). The latter is not exclusive to the hetero-chalcogen case, and would mainly be visible in a modification of the phonon frequencies (LO/TO splitting) compared to the bulk. The main effect is therefore the static charge redistribution, polarizing the two layers in opposite ways, and leading to a net compression of the interlayer spacing. The (dispersive) vdW interaction between differing chemical species is not expected to yield this kind of anomaly, and we confirm numerically that the S-Se interaction is simply the average of the S-S and Se-Se (by construction in the Grimme D3 scheme).

A number of recent works (e.g. References~\citenum{idrees2019} and~\citenum{Zhang2020}) have examined Janus TMD structures, where the chalcogen on one side of a layer has been substituted. These can be made experimentally~\cite{Zhang2020} by exposing (e.g.) \mos~to a hydrogen plasma and Se vapor. If there are islands or flakes of \mos~on top of a \mos~surface, the S against S interface is preserved, and the Janus flake now presents an intrinsic dipole. Epitaxy is enforced by the synthesis process. The interlayer distance is also reduced by this arrangement, as in our calculations. The authors do not specify the mechanism for the layer attraction, but it seems natural to expect it is electrostatic as in our work, and not simply vdW interactions. In Ref.~\citenum{idrees2019} bilayers of Janus TMDs are studied theoretically and again the intrinsic dipole leads to an electrostatic attraction/repulsion between the layers: head to tail dipoles yield a smaller interlayer spacing. Again the origin of the contraction observed here and in Janus bilayers is the same. 

\section{Conclusions}\label{sec: conclusions}

In summary, we have calculated the structural, electronic, and vibrational properties of ideal commensurately-stacked heterostructures containing \mos, \mose, \ws, \wse, \tis, \tise, \zrs, and \zrse~using the \abinit~software package. For dissimilar chalcogen atoms (group II), we find there is a strong contraction of the c lattice parameter, localized in the van der Waals gap distances between the layers. We trace this contraction back to an induced electrostatic effect and a polarization of the layers with a partial spatial charge transfer, leading to a net attraction between the layers. This effect is disentangled from the artificially imposed epitaxy, by comparing with an intermediate bulk structure, which is constrained in-plane to the heterostructure lattice constant (and relaxed out of plane). For identical chalcogens (group I) there is only a small compression, which is mainly due to the epitaxy.
 

Calculations of the dynamical stability of these compounds, via the inter-atomic force constants and the phonon spectra, indicate that the semiconducting heterostructures are stable. In the hexagonal heterostructures there is a clear separation of the optical vibrational modes for each layer, indicating a weaker inter-layer interaction in these materials. The acoustic modes mix more, especially in the hetero-chalcogen group II cases. In the trigonal heterostructures, we find that the charge density wave instability from the Ti based compounds is affected strongly: Counter-intuitively, when one generates a heterostructure of \tis~and \tise, the resulting compound is dynamically stable due to a small charge transfer between the layers.
The \tiszrs~combination is semiconducting with a small gap, but dynamically unstable, with an unstable polar mode at $\Gamma$.
\tisezrse~is semi-metallic, and retains the CDW instability at the M point of the Brillouin zone. These three evolutions are linked to interlayer charge transfers and not only strain, and may be observable in experimental bilayers. Ferroelectricity in~\tiszrs~is a particularly exciting possibility.

While these commensurate heterostructures will not all be the experimental ground state (preferring to twist and create moir\'e patterns), they provide insight into heterostructure property modifications, and guidance to understand how the individual layers change due to stacking in heterostructures. We note several effects which should be measurable: interlayer distances, phonon frequency shifts, stabilization of charge density waves (\tistise), and the appearance of ferroelectricity in \tiszrs. The proximity and layer polarization should also have an impact on transport through the layers.

\begin{acknowledgments}
N.A.P. and M.J.V. gratefully acknowledge funding from the Belgian Fonds National de la Recherche Scientifique (FNRS) under grants PDR T.1077.15-1/7 and  T.0103.19-ALPS. M.J.V. acknowledges a FNRS sabbatical ``OUT'' grant at ICN2 Barcelona, as well as ULi{\`e}ge and the F\'ed\'eration Wallonie-Bruxelles (ARC AIMED G.A. 15/19-09). 
Computational resources have been provided by the Consortium des Equipements de Calcul Intensif (CECI), funded by FRS-FNRS G.A. 2.5020.11; the Zenobe Tier-1 supercomputer funded by the Gouvernement Wallon G.A. 1117545; and by a PRACE-3IP DECI grants 2DSpin and Pylight on Beskow, and RemEPI on Archer (G.A. 653838 of H2020). 
This publication is based upon work of the MELODICA project, funded by the EU FLAG-ERA\_JTC2017 call.
N.A.P. and A.D. contributed equally to this work.
\end{acknowledgments}
\bibliography{TMD_heterostructures}

\end{document}